\documentclass[aps,twocolumn,prd,superscriptaddress]{revtex4-2}

\usepackage[normalem]{ulem}

\usepackage{diagbox}
\usepackage{mathtools}

\usepackage{amsfonts}
\usepackage{mathrsfs}
\usepackage{bbm}
\usepackage{slashed}
\usepackage{dsfont}
\usepackage{xspace}
\usepackage{siunitx}
\usepackage{hyperref}
\usepackage[nameinlink]{cleveref}
\usepackage{appendix}
\usepackage{graphicx}
\usepackage{color}
\usepackage{array}

\usepackage{blindtext}
\usepackage{placeins}
\usepackage{booktabs}
\usepackage{makecell}
\usepackage{epstopdf}
\usepackage[caption=false]{subfig}
\usepackage{xspace}
\usepackage{siunitx}
\usepackage{hyperref}
\usepackage[nameinlink]{cleveref}
\usepackage{appendix}

\usepackage{xifthen}
\usepackage{xcolor}
\hypersetup{
	colorlinks,
	linkcolor={red!75!black},
	citecolor={blue!75!black},
	urlcolor={blue!75!black}
}

\setkeys{Gin}{width=0.48\textwidth}

\graphicspath{
	{./figures/}
	{../figures/}
}

\def\s0#1#2{\mbox{\small{$ \frac{#1}{#2} $}}}
\def\0#1#2{\frac{#1}{#2}}

\newcommand{\Tr}{{\text{Tr}}}

\newcommand{\imag}{\text{i}}

\usepackage{multirow}
\usepackage{colortbl}
\definecolor{kugray5}{RGB}{224,224,224}
\newcommand{\PreserveBackslash}[1]{\let\temp=\\#1\let\\=\temp}
\newcolumntype{C}[1]{>{\PreserveBackslash\centering}p{#1}}
\newcolumntype{R}[1]{>{\PreserveBackslash\raggedleft}p{#1}}
\newcolumntype{L}[1]{>{\PreserveBackslash\raggedright}p{#1}}

\begin{document}

\title{Flavor dependence  of chiral symmetry breaking and the conformal window}

\author{Yi-huai Chen}
\email{3120231486@bit.edu.cn}
\affiliation{School of Physics, Beijing Institute of Technology, 100081 Beijing, China}

\author{Yi Lu}
\email{qwertylou@pku.edu.cn}
\affiliation{Department of Physics and State Key Laboratory of Nuclear Physics and Technology, Peking University, Beijing 100871, China}
\affiliation{Center for High Energy Physics, Peking University, 100871 Beijing, China}
\affiliation{Collaborative Innovation Center of Quantum Matter, Beijing 100871, China}

\author{Zhi-wei Wang}
\email{zhiwei.wang@uestc.edu.cn}
\affiliation{School of Physics, The University of Electronic
Science and Technology of China, Chengdu, China}

\author{Yu-xin Liu}
\email{yxliu@pku.edu.cn}
\affiliation{Department of Physics and State Key Laboratory of Nuclear Physics and Technology, Peking University, Beijing 100871, China}
\affiliation{Center for High Energy Physics, Peking University, 100871 Beijing, China}
\affiliation{Collaborative Innovation Center of Quantum Matter, Beijing 100871, China}

\author{Fei Gao}
\email[]{fei.gao@bit.edu.cn}
\affiliation{School of Physics, Beijing Institute of Technology, 100081 Beijing, China}

\begin{abstract}

We investigate the phase structure of Quantum Chromodynamics (QCD)  in the vacuum as a function of quark flavor number $N_f$ within the chiral limit. 
By self-consistently solving the coupled DSEs for the quark and gluon propagators in a minimal QCD scheme, 
we elucidate the nonperturbative dynamics governing dynamical chiral symmetry breaking. Our calculations determine a critical flavor number of 
$N_f^c=6.81$ which marks the chiral symmetry restoration of quarks.  Further  analysis reveals the critical exponents of the chiral condensate as $ -\langle\bar{\psi} {\psi}\rangle\sim |N_f-N_f^c|^{0.53(9)}$, characterized the second order feature of this phase transition of chiral symmetry. Additionally, we discuss the implications for  the walking regime towards the conformal window at larger flavor. 
\end{abstract}

\maketitle

\section{Introduction}
Quantum Chromodynamics (QCD) which governs the strong interaction, exhibits profoundly different behaviors across energy scales. While perturbation theory successfully describes high-energy processes via asymptotic freedom, 
the low-energy regime remains challenging due to nonperturbative properties in association to confinement, dynamical chiral symmetry breaking (DCSB). The dependence of the quark flavor $N_f$ in QCD is therefore crucial to illustrate the transition of the dynamics  between the perturbative and non perturbative region. As has been widely investigated,  the beta function for the coupling becomes positive for $N_f>16.5$ with $N_c=3$ in QCD.  Just below this critical flavor number where QCD is no longer asymptotic free, an infrared fixed point is also found, i.e.,   the  Caswell-Banks-Zaks (CBZ) fixed point~\cite{Caswell:1974gg,Banks:1981nn}.  Consequently, there exists the conformal window between $N_f=16.5$ and the critical flavor number of CBZ fixed point   for which the theory is infrared conformal~\cite{Caswell:1974gg,Banks:1981nn,Dietrich:2006cm,Braun:2011pp,Rychkov:2016iqz,Hansen:2017pwe}. Moreover, below this conformal window, it  may result in a slowly running,  i.e.,  the walking regime,  disctinct from the situation for lower flavor number which there typically exist the chiral symmetry breaking and confinement~\cite{Lane:1991qh,Hopfer:2014zna,Hasenfratz:2022qan,Goertz:2024dnz}.

While the upper boundary of the conformal window at $N_f=16.5$ can be accessed through perturbative methods, determining its lower boundary which is signaled by the  infrared  CBZ fixed point  demands non-perturbative analysis~\cite{Miransky:1996pd,Gies:2005as,Armoni:2009jn}. The precise determination of  this lower boundary remains  challenging as the various methods deliver different predictions~\cite{Braun:2009ns,Pica:2010xq,Hopfer:2014zna,Bashir:2013zha,Lee:2020ihn,Goertz:2024dnz}.  The Lattice QCD simulations have confirmed the conformality of $N_f=12$~\cite{Appelquist:2007hu,Deuzeman:2009mh,Hasenfratz:2011xn,Aoki:2012yd}, and found that  the case of $N_f=8$ shows  the feature that the system is  in the pre conformal walking regime~\cite{Miura:2012zqa,Aoki:2013xza,LatKMI:2016xxi,Hasenfratz:2022qan}.  Notably,  in the walking regime, it has been argued that this might result from a symmetric mass generation  without the chiral symmetry breaking~\cite{Hasenfratz:2022qan}, and hence, it is of great importance to conduct an anatomic  analysis on the  quark and gluon dynamics separately to really understand the walking regime. 

The functional QCD approaches including  the Dyson-Schwinger equations (DSE)  approach~\cite{Hopfer:2014zna,Bashir:2013zha}  and also the functional renormalization group (fRG) approach~\cite{Goertz:2024dnz} opens an access to give such a comprehensive analysis on the flavor dependence of QCD. As the result of  quark propagator straightforwardly gives the  scale of chiral symmetry breaking, while the gluon propagator can define the mass scale of gluon,  potentially  originating the symmetric mass generation. Here in specific, we apply the DSEs approach which offers a self-consistent framework for investigating conformal dynamics via the infrared behavior of propagators,  revealing the emergence of the dynamical mass gap  of quark and gluon. 
We perform a  non-perturbative analysis  by solving the coupled equations of quark and gluon propagator's Dyson-Schwinger equations in the up to date minimal QCD scheme which quantitatively captures the correct  running behavior of the quark,  gluon propagator and also quark gluon vertex~\cite{Gao:2020qsj,Lu:2023mkn,Lu:2025cls}.  Such a  quantitative calculation helps to classify the actual scenario that happens in the phase transition with  varying the quark flavor and also at the finite temperature.

In \Cref{sec:minimalDSE} we present the minimal QCD scheme of Dyson Schwinger equations in particular for the extension to arbitrary flavor number.   In \Cref{sec:numerics} we first check the consistency of the results  at $N_f=2$ and $N_f=3$  and then give the results for the critical flavor number where the chiral symmetry is restored. We also discuss the results in association to the relation between this critical point and the conformal window.  In \Cref{sec:Summary}, we briefly summarize the main results and provide an outlook on the next steps. 

%%%%%%%%%%%%%%%%%%%%%%%%%%%%%%%%%
\section{Framework of  Dyson-Schwinger Equations}
\label{sec:minimalDSE}
 Central to our framework is to solve the coupled DSEs for quark and gluon propagator. The quark gap equation serves as the cornerstone in the framework which  determines the running of quark masses and order parameter of chiral symmetry, requiring only the gluon propagator and quark-gluon vertex as input.
We implement a minimal parametrization of quark gluon vertex based on the Slavnov-Taylor identity and the requirement of renormalization. As demonstrated in Refs~\cite{Gao:2020qsj,Lu:2023mkn,Lu:2025cls}, this treatment yields results consistent with  more elaborate calculations that explicitly solve quark gluon vertex DSE~\cite{Gao:2020fbl,Gao:2021wun}.  For the gluon propagator, we adopt the methodology introduced in Ref.~\cite{Fischer:2012vc,Fischer:2014ata} using the quenched result for $N_f=0$ obtained from lattice results~\cite{Sternbeck:2005tk} as a baseline, and calculating the quark loop in the self energy of gluon gap equation.  One may also expand the calculation on the gluon propagator at $N_f=3$ which is also widely studied in functional approaches and lattice QCD simulations. The consistency between these different expansion schemes  	will be quantitatively demonstrated in the next section.

We begin by briefly deriving the DSEs for QCD from the path integral formalism. The generating functional of the theory is given by:
\begin{equation}
\begin{split} 
 Z[J,\eta,\bar{\eta}]= \int{\cal D}A{\cal D}q{\cal D}\bar{q}\,e^{\imag S+\int JA+\bar{\eta}q+\eta \bar{q}}
\label{eq:functional}
\end{split}
\end{equation}

where $S$ is the action of QCD:
\begin{equation}
\begin{split} 
 S[\bar{q},q,A_\mu]= \int d^4x [-\frac{1}{4}F^a_{\mu\nu}F^{a\mu\nu}+\bar{q}(i\gamma^\mu D_\mu -m)q]
\label{eq:action}
\end{split}
\end{equation}

The Lagrangian involves the gauge covariant derivative and the non-Abelian field strength tensor $F^a_{\mu\nu}$. To quantize the theory, we employ the Faddeev-Popov procedure in its non-Abelian formulation to fix the gauge. The Becchi-Rouet-Stora (BRS) symmetry of the gauge-fixed action leads to the Slavnov-Taylor identities (STIs)—the non-Abelian generalizations of the Ward-Takahashi identities (WTIs) in QED. This gauge-fixing process necessitates the introduction of Faddeev-Popov ghost fields to ensure that physical observables remain gauge-independent.
The generating functional of the theory can be Legendre-transformed to introduce the quantum effective action $\Gamma$, whose functional derivatives are the one-particle irreducible (1PI) correlation functions\cite{Alkofer:2000wg,Dupuis:2020fhh}:
\begin{equation}
\begin{split} 
\Gamma[\phi^c]=-\ln Z[j]+\phi^c_i j_i \,\,\, ,\,\,
\phi^i_c=\frac{\delta \ln Z[j]}{\delta j_i}
\end{split}
\label{eq:1pi}
\end{equation}
Based on the action principle, the Green's functions of QCD can be derived by performing functional variations on the generating functional, such as the quark propagator:
\begin{equation}
\begin{split} 
S[x,y,A_\mu]=\left({\frac{\delta^2\Gamma}{\delta \bar{q}(x)\delta q(y)}}\right)^{-1}_{\bar{q}=q=0}
\end{split}
\label{eq:qu Green funct}
\end{equation}

\begin{figure}[t]
\centering
\includegraphics[width=0.9\columnwidth]{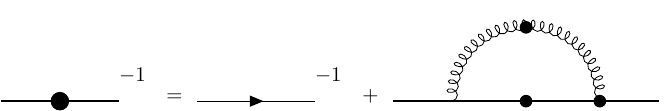} 
\caption{The Dyson Schwinger equation for quark propagator, i.e., the quark gap equation.}
\label{fig:quarkDSE}
\end{figure}

The DSE for quark propagator, i.e. the quark gap equation, can be expressed in momentum space (cf.~\cref{fig:quarkDSE}) as:

\begin{equation}
\begin{split} 
& S^{-1}\left(p\right)=S_{0}^{-1}\left({p}\right)+\Sigma(p),\\
\end{split}
\label{eq:quarkDSE}
\end{equation}
with
\begin{equation}
\Sigma(p)=\frac{4}{3}g^2\int \frac{d^{4}q}{(2\pi)^4} D_{\mu\nu}\left(p-q\right)\gamma_{\mu}S\left(q\right)\Gamma_{\nu}\left(q,p\right). \notag
\label{eq:quarkself}
\end{equation}

The dressed quark propagator $S(p)$ is obtained by solving the Dyson-Schwinger equation (DSE) depicted in \cref{fig:quarkDSE}, which involves the bare quark propagator $S_{0}(p)$ and the quark self-energy $\Sigma(p)$. The interaction kernel for the self-energy is defined by the gluon propagator 
$D_{\mu\nu}(k)$ and the quark-gluon vertex $\Gamma_{\nu}(p,q)$. The DSE of the gluon propagator  is also employed here for the calculation, but only the self energy of quark loop is involved. The quark gluon vertex  $\Gamma_{\nu}(p,q)$ is determined by the combination of quark and gluon dressing functions. This self-consistent interplay between the quark propagator and the interaction kernel is fundamental to the DSE formalism and drives dynamical chiral symmetry breaking (DCSB).  The key components and justification for the quark gap equation are detailed in Section~\ref{sec:minimal}, while for the calculation of the gluon propagator $D_{\mu\nu}(k)$ is put in Section~\ref{sec:quarkloop}.

%%%%%%%%%%%%%%%%%%%%
\subsection{The quark gap equation}
\label{sec:minimal}

 To solve the non perturbative gap equation, one can expand the quark and gluon propagator  on a complete basis with the Dirac vector and scalar dressing functions, 
$A(p^2)$ and $B(p^2)$, respectively. In Landau gauge, with the color indices being factored out  directly, the quark and gluon propagators take the following form:
\begin{equation}
	S^{-1}\left(p\right)=\imag \slashed{p}A(p)+B(p),
	\label{quarkpropa}
\end{equation}
and 
\begin{align}
  D_{\mu\nu}(k) =G_A(k)\Pi^\bot_{\mu\nu}(k) \,, \quad
  \Pi^\bot_{\mu\nu}(k) = \delta_{\mu\nu} - \frac{k_{\mu} k_{\nu}}{k^2}\,,
 \label{eq:gluon}
\end{align}
with
\begin{equation}
  G_A(k)=Z(k)/k^2,
 \label{eq:gluondressing}
\end{equation}
where $Z(k^2)$ is the gluon dressing function.  
 {In Landau gauge the gluon propagator is transverse, $k_\mu D_{\mu\nu}(k)=0$, so only transverse quark--gluon vertex components contribute to $\Sigma(p)$; longitudinal parts are projected out by $\Pi^{\perp}_{\mu\nu}(k)$.
}
The last key component is the quark-gluon vertex. In the vacuum, its complete tensor basis consists of twelve independent structures, eight of which comprise the transverse part which is required in the Landau gauge:

 \begin{equation}
	\Gamma_{\mu}(q,p)=\sum_{i=1}^{12}{\cal T}_{\mu}^{i}(q,p)\lambda^i(q,p).
 \end{equation}

 {Continuum and lattice studies further indicate that the leading nonperturbative strength
is carried by the Dirac (vector) and Pauli (tensor) structures over a wide momentum range~\cite{Qin:2013mta,Gao:2021wun,Chang:2021vvx}.}
The quark gluon vertex can be then simplified as:
 \begin{equation}
	\Gamma_{\mu}(q,p)={\cal T}_{\mu}^{1}(q,p)\lambda^1(q,p)+{\cal T}_{\mu}^{4}(q,p)\lambda^4(q,p),
 \end{equation}
 with 
\begin{align}\nonumber
{\cal T}_\mu^{(1)}(q,p) = &\,  -\imag\gamma_\mu \,,\\[1ex]\nonumber 	
		{\mathcal{T}_\mu^{(4)}}(q,p)
	= &\,  -\imag\sigma_{\mu\nu} k^{\nu}\, \,,\qquad \sigma_{\mu\nu} = \frac{\imag}{2} \, [\gamma_{\mu},\gamma_{\nu}]\,.
\label{eq:vertsx147}
\end{align}
Where $k$ is the gluon propagator momentum.
 The dressing of the tensor structures is constrained by the STI and the requirement of renormalization.
 {Since the STI constrains the contracted vertex $k^\mu\Gamma_\mu$ and
$k^\mu\sigma_{\mu\nu}k^\nu=0$, the ghost factor appearing in the STI can only dress the
Dirac component in our two-term truncation, which leads to
$\lambda^1(q,p)=F(k^2)\Sigma_A(q,p)$ (the remaining STI-induced structures are omitted here).
}
Thus we have:
 \begin{equation}
	\lambda^1(q,p)=F(k^2)\Sigma_A(q,p) \, ,\quad
	\Sigma_A(q,p)=\frac{A(p)+A(q)}{2}.
 \end{equation}
 Where $F(k^{2})$ is the ghost dressing $k^2D(k^2)$, $D(k^2)$ is the ghost propagator. 
 \begin{equation}
\lambda^4(q,p)=\Delta_B(q,p)/\sqrt{Z(k^2)} \, ,\quad
	\Delta_B(q,p)=\frac{B(p)-B(q)}{p^2-q^2}.
\end{equation}

For the ghost dressing function that is required in  the $\lambda^1$ expansion, the increase  of the flavor number produces only subleading corrections to the ghost propagator, as it does not directly appear in the self energy of the ghost propagator and comes only from the quark loop of the gluon propagator.  We therefore neglect its flavor dependence and adopt the result  that is   applied  in the physical parameter space~\cite{Lu:2023mkn,Zheng:2023tbv,Lu:2025cls}.  In the following, we will elaborate how to include the flavor dependence of the gluon propagator.

%%%%%%%%%%%%%%%%%%%%%%%%%%%%%

\begin{figure}[t]
\centering
\includegraphics[width=0.9\columnwidth]{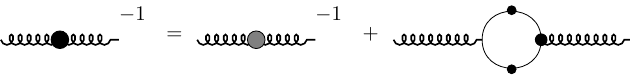} 
\caption{The Dyson-Schwinger equation for the gluon propagator, with the grey point standing for the quenched gluon propagator calculated in Yang-Mills theory and thus only the quark loop is left aside in this DSE of gluon propagator. }
\label{fig:gluonDSE}
\end{figure}

\subsection{The quark loop in gluon DSE }
\label{sec:quarkloop} 

The gluon propagator {satisfies the Dyson--Schwinger equation}
\begin{equation}
	D_{\mu\nu}(k)^{-1}=[{D_0(k)}]_{\mu\nu}^{-1}+\Pi_{\mu\nu}(p)^{gh,gl,qu}, 
	\label{gluonDSE}
\end{equation}
{where $\Pi_{\mu\nu}^{gh}$, $\Pi_{\mu\nu}^{gl}$, and $\Pi_{\mu\nu}^{qu}$ denote the ghost-, gluon-, and quark-loop
contributions to the gluon self-energy, respectively.}   The Yang-Mills sector is relatively separable and has been widely studied~\cite{Sternbeck:2005tk,Bogolubsky:2009dc,Oliveira:2012eh,Aguilar:2012rz,Cyrol:2016tym,Huber:2018ned,Eichmann:2021zuv,Aguilar:2021uwa,Ferreira:2025anh}. {In the present setup, the contributions of gluons and ghost loop are absorbed into the quenched gluon propagator introduced above (grey blob in Fig.~\ref{fig:gluonDSE}),
and we explicitly compute the quark-loop back-reaction $\Pi_{\mu\nu}^{qu}$ below.}

The quark loop's contribution:
\begin{equation}
	\Pi_{\mu\nu}^{\textcolor{blue}{qu}}(p)=\frac{-g^2N_{f}}{2}\int \frac{d^4q}{(2\pi)^4} \, \Tr{[\gamma_{\mu}S(q)\Gamma_{\nu}(q,p)S(k)]},
\end{equation}
where $p$ is the external gluon momentum {while} $q$ and $k=p-q$ denote the internal quark loop momenta.
We adopt the transverse tensor structures $\lambda_1$ and $\lambda_4$ for the quark-gluon vertex, consistent with the gap equation. 
{The Pauli dressing \(\lambda_4\) is driven by dynamical chiral symmetry breaking: in our STI-inspired
construction \(\lambda_4\propto \Delta_B\) and therefore becomes suppressed when the quark mass function
\(M(p^2)=B(p^2)/A(p^2)\) decreases at large \(N_f\) toward the (near-)conformal regime.} 
In fact,  the numerical result confirms that $\lambda_4$ term  does not have impacts yet at  small flavor number and can be safely neglected.

In the gluon DSE, the introduction of an ultraviolet cutoff $\Lambda$ induces spurious quadratic divergences. To properly extract the momentum-dependent gluon dressing function in a gauge-invariant manner, we employ the Brown-Pennington projector \cite{Brown:1988bm,Fischer:2003rp}, which  is defined as:
\begin{equation}
	{\cal P}_{\mu\nu}(p)=\delta_{\mu\nu}-\frac{4p_{\mu}p_{\nu}}{p^2} \,.
\end{equation}
 {Note that $P_{\mu\mu}(p)=0$, i.e. $P_{\mu\nu}(p)$ is traceless, and therefore removes any pure-trace contribution
$\propto \delta_{\mu\nu}$ (including the spurious $\sim \delta_{\mu\nu}\Lambda^2$ term induced by a hard cutoff $\Lambda$).}
Contracting  the Brown-Pennington projector with the free Lorentz indices of the quark loop, and one gets:
\begin{eqnarray}
	\Pi(p)=&&{\cal P}_{\mu\nu}\Pi^q_{\mu\nu}\\
	=&&-\frac{4}{3}g^2N_{f}\int \frac{d^4q}{(2\pi)^4} \lambda_{1}A(k)A(q)(\frac{4\,kp\,kq}{p^2}-kq)\,.\notag
\end{eqnarray}
{In Landau gauge, this contraction isolates the transverse (physical) part of the quark polarization tensor that enters the gluon dressing function.
After performing the Dirac trace, the scalar dressing $B$ contributes only to the pure-trace part
$\propto \delta_{\mu\nu}$ (while the mixed $AB$ terms vanish due to an odd-$\gamma$ trace), which is removed by the
traceless Brown--Pennington projector $P_{\mu\nu}$; this is why Eq.~(16) depends only on $A(q^2)A(k^2)$.
}
Note that a complete dynamical description must account for the back-reaction of quark loops on the gluon propagator. Contributions from quark loops of different flavors collectively determine the dynamics of the gluon propagator. In particular, increasing the number of flavors enhances screening, which dynamically suppresses the infrared strength of the gluon propagator and, in turn, weakens the effective quark-gluon coupling.

This completes the minimal QCD scheme of DSEs with the extension for  large flavor numbers.  {Note, however, that this quenched-baseline subtraction (difference) scheme, in which the quark loop is inserted into the quenched gluon propagator, should not be pushed to very large $N_f$}
In particular, the fact that it always maintains a mass gap in the gluon propagator indicates that it cannot correctly  {capture} the dynamics {once} the theory loses the asymptotic freedom  {at} $N_f\geq 16.5$, {where large-$N_f$ studies suggest a qualitatively different (possibly UV-safe) RG behavior}~\cite{Mann:2017wzh,Antipin:2017ebo,Antipin:2018zdg}.

%%%%%%%%%%%%%%%%%%%%
\section{numerical results}
\label{sec:numerics}

We present our numerical results for theories with different numbers of flavors $N_f$. Our computation starts from the chiral-limit quark mass function for $N_f=0$, obtained from the gap equation. For each added flavor, we compute the quark loop contribution to the gluon propagator and self-consistently update the quark gap equation. This process is iterated until full convergence is achieved across all flavors. 
\begin{figure}[t]
\centering
\includegraphics[width=0.9\columnwidth]{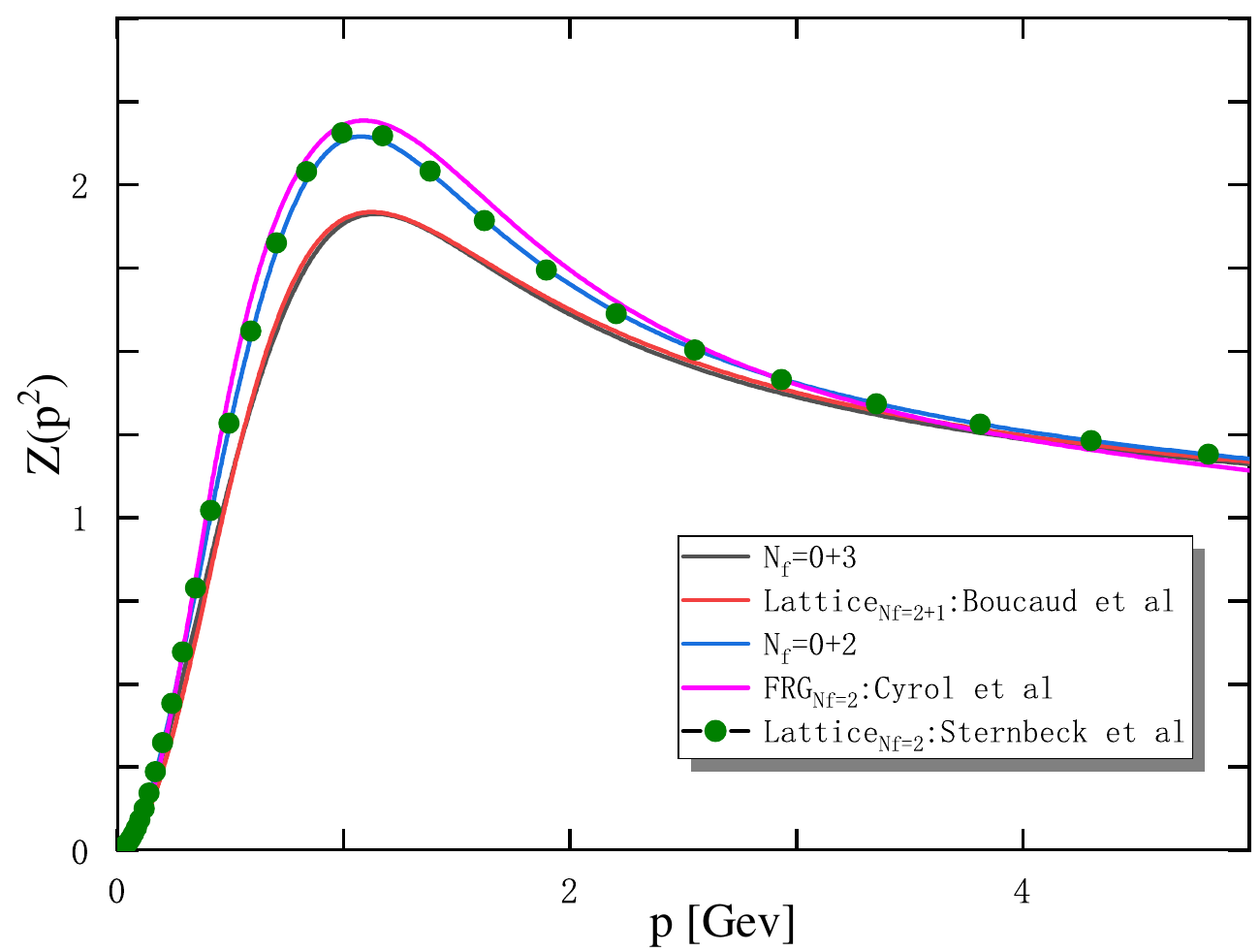} 
\caption{Gluon propagator dressing function in the current work at $N_f=0+2$ and $0+3$,  in comparison to the previous results of  $N_f=2$ flavor~\cite{Sternbeck:2005tk,Cyrol:2017ewj} and $2+1$ flavor~\cite{Boucaud:2018xup,Gao:2021wun}.   }
\label{fig:dressing compare}
\end{figure}

\begin{figure}[t]
\centering
\includegraphics[width=0.9\columnwidth]{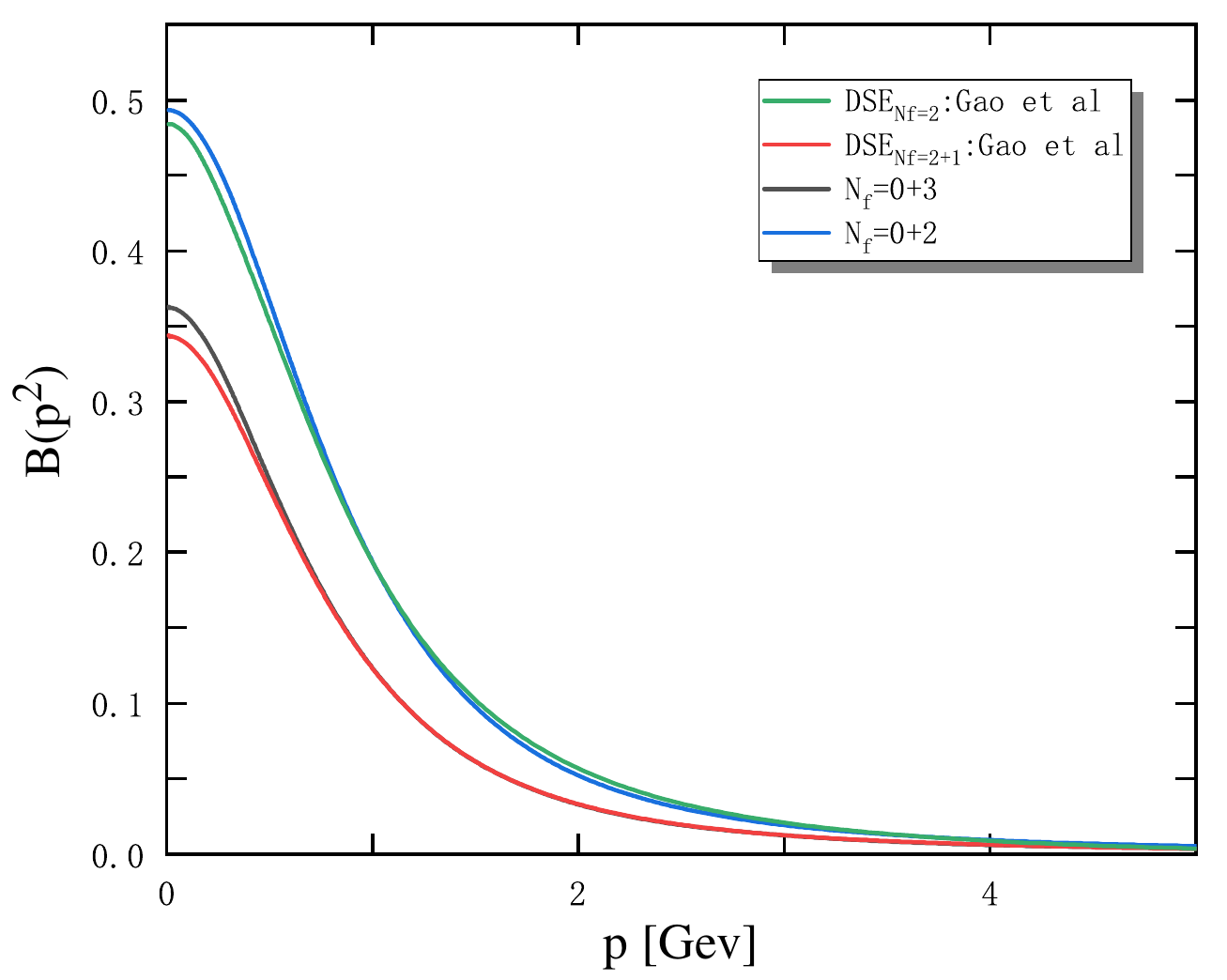} 
\caption{The obtained scalar  function  of quark propagator which determines the strength of the chiral symmetry breaking, in comparison with the previous results at $N_f=$ 2 and 2+1 flavors~\cite{Gao:2020qsj}. }
\label{fig:mass compare}
\end{figure}

\label{sec:numericalresults}
\subsection{The benchmark calculation  for the gluon and quark propagator}
Our calculation starts from the case  at $N_f=0$ with the   quenched gluon propagator from lattice QCD being put in  and we rigorously solve the Dyson–Schwinger equation (gap equation) to obtain the quark propagator in the minimal QCD scheme as described above. To make sure our calculation captures the correct quark number dependence, we first verify  the consistency between the resulting $N_f=2$ and $3$ system and the  previous benchmark results of  $N_f=2$ and $N_f=2+1$   from lattice QCD simulation and functional QCD approaches.

We first depict the gluon dressing function for $N_f=0+2$ and $N_f=0+3$ {(with $N_f=0+N$ denoting a quenched-gluon baseline plus the $N$-flavor quark-loop backreaction),} together with the direct calculation results from lattice QCD and functional QCD approach in ~\cref{fig:dressing compare}.  Our results are in good agreement  with the  previous results from direct calculation, confirming that the difference scheme that inserts the quark loop into the quenched gluon propagator is valid for a large range of flavor number. In particular, due to the consistency of the results at $N_f=3$, {one may equivalently take the $N_f=3$ result as the baseline and extend to larger $N_f$ by adding the extra-flavor quark-loop backreaction}, which further strengthens the reliability of the scheme for large flavor number. 

The   scalar  function $B(p^2)$ of quark propagator for  $N_f=0+2$ and $N_f=0+3$ are presented in \cref{fig:mass compare}, which also show good consistency with the {previous direct functional-QCD result of Ref.~\cite{Gao:2020qsj}, where the DSE calculation is assisted by fRG input at $N_f=2$, i.e.\ treating unquenching effects self-consistently rather than using a quenched-gluon baseline with an inserted quark-loop backreaction as in our $N_f=0+N$ scheme}. The wave function of quark propagator $A(p^2)$ has $5\,\%$ deviation in comparison to the previous results, due to the neglect of   the other Lorentz structures  in the quark gluon vertex. Nevertheless, the scalar function of the quark propagator is the central element as the criterion of the chiral symmetry breaking. The consistency  of the scalar function of the quark propagator at $N_f=0+2$ and $N_f=0+3$  validate  the scheme for  the extension to large flavor number.

\begin{figure}[t]
\centering
\includegraphics[width=0.9\columnwidth]{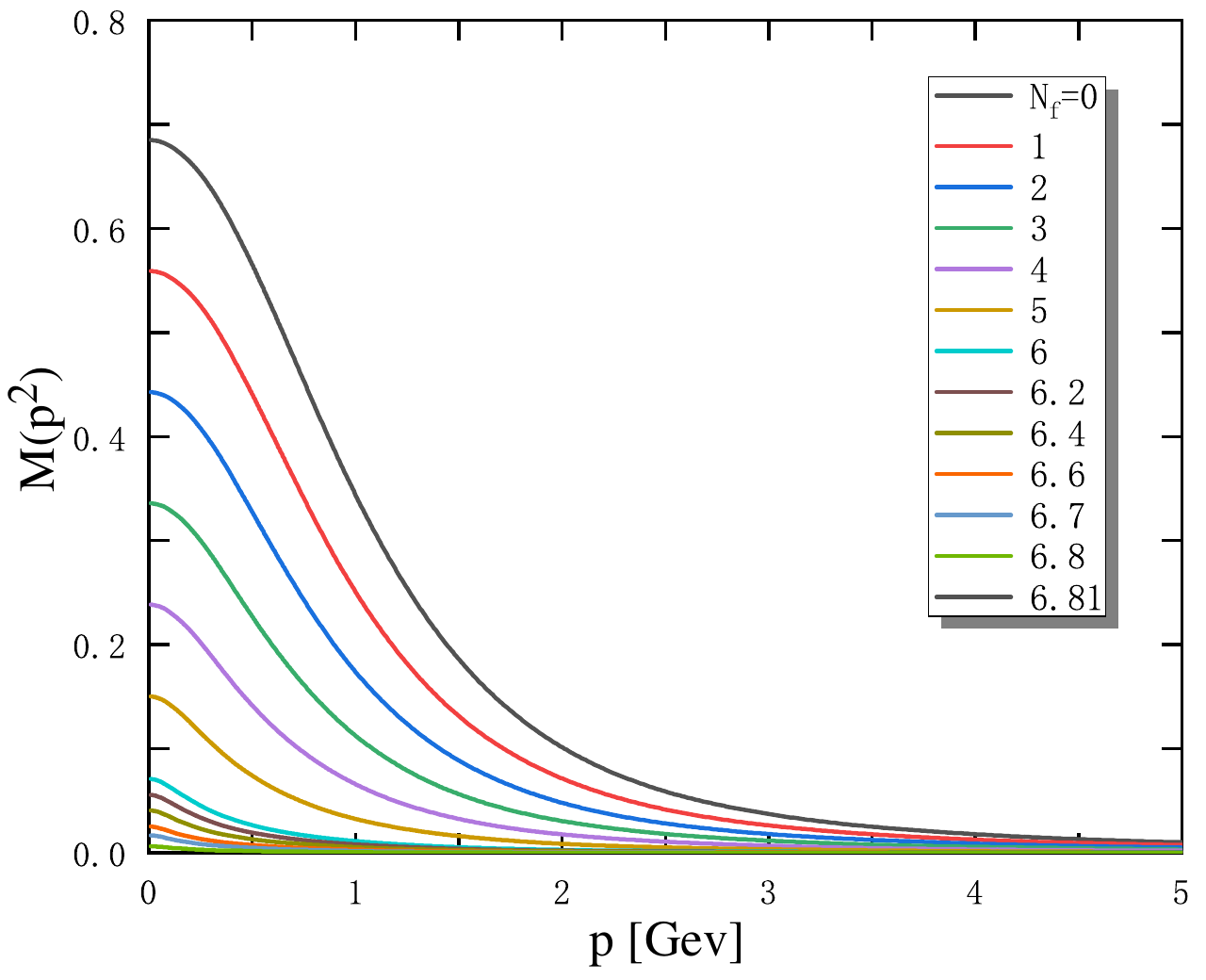} 
\caption{The evolution of quark mass function along with the change of flavor numbers. At $N_f^c=6.81$, the quark mass function vanishes.}
\label{fig:quarkmass}
\end{figure}
\subsection{The critical  flavor number for chiral symmetry restoration and conformal window}

After solving the coupled DSE equations, one can study the chiral symmetry directly from the quark propagator. As illustrated in \cref{fig:quarkmass}, the quark mass function $M(p^2)=B(p^2)/A(p^2)$ decreases monotonously as the flavor number increases, representing the restoration of the chiral symmetry.  At a critical flavor number $N_f^c=6.81$, the mass function vanishes and the  dynamical mass generation ceases entirely.

\begin{figure}[t]
\centering
\includegraphics[width=0.9\columnwidth]{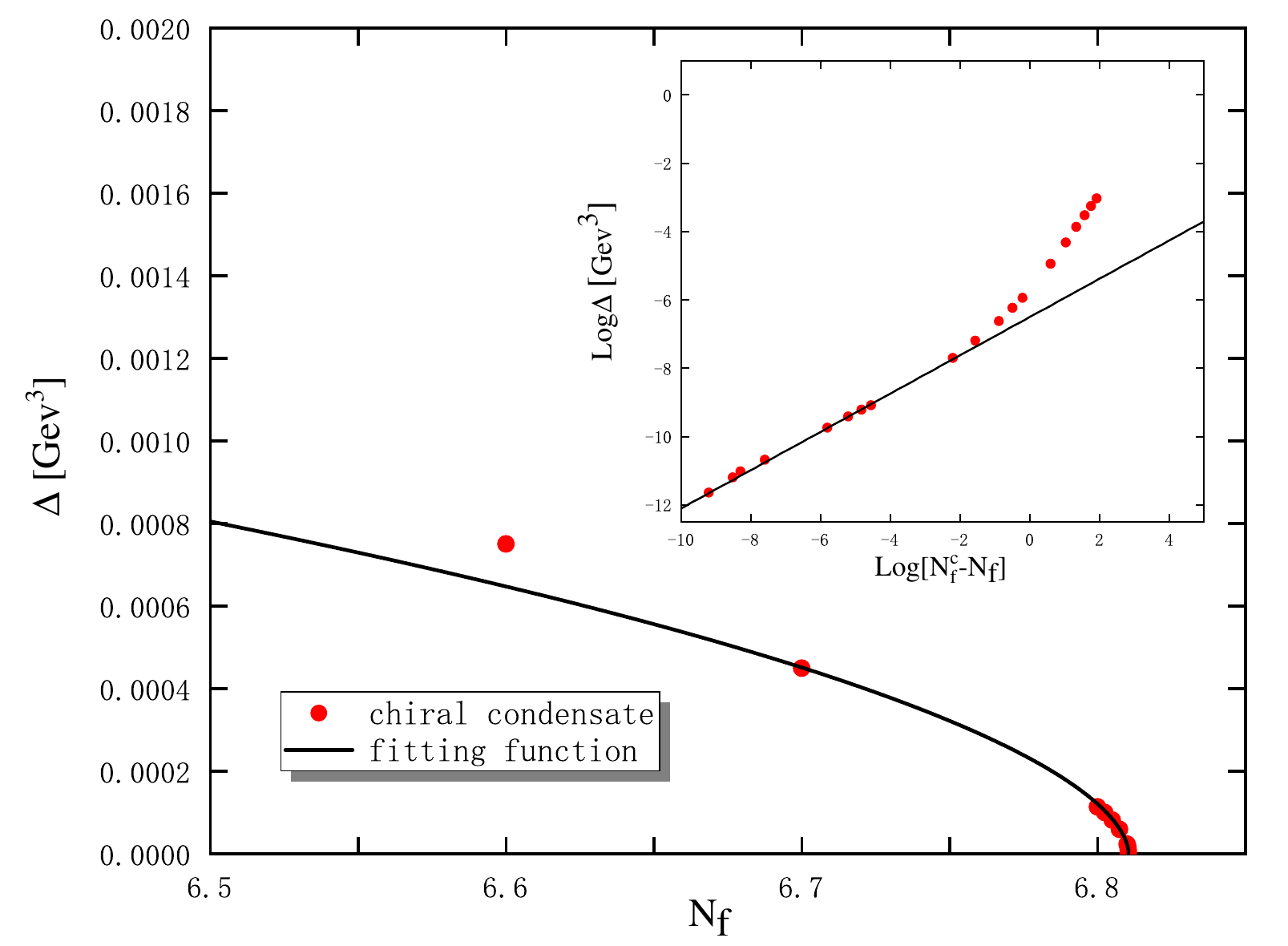} 
\caption{The Chiral condensate $\Delta$ as a function of flavor number $N_f$. Near the critical flavor $N_f^c=6.81$ where the chiral condensate is zero, it can be fitted by a power law behavior as $|N_f-N_f^c|^{0.53(9)}$.}
\label{fig:chiral condensate}
\end{figure}
As usual, we apply  the chiral condensate as the order parameter of the chiral symmetry. In chiral limit without the current quark mass, no quadratic divergence is present and the direct trace of the quark propagator gives the correct result up to the renormalization.  One  can directly define the chiral condensate $\Delta$ as:

\begin{equation}
{{\Delta \equiv}} -\langle\bar{\psi}\psi\rangle = \int \frac{d^4p}{(2\pi)^4} {\rm Tr} [ S(p)],
\end{equation}
where  the trace includes  the color and Dirac indices.  For $N_f=3$, the chiral condensate at $\mu=2$ GeV is $$-\langle\bar{\psi}\psi\rangle^{\frac{1}{3}}=274\, \rm{MeV}, $$  which is consistent with the results from lattice QCD and also functional QCD results~\cite{FlavourLatticeAveragingGroup:2019iem, Gao:2021wun}.  

Further quantitative evidence for the chiral condensate is provided in \cref{fig:chiral condensate}, which demonstrates a characteristic flavor-induced suppression of the chiral condensate.  Moreover, we find that near the critical flavor number, the condensate shows scaling behavior, as it can be described by a power law behavior as:

\begin{equation}
\Delta\sim |N_f-N_f^c|^{\frac{1}{\Theta}}\, ,\Theta=1.88\pm0.032.
\end{equation}
 In the logarithmic plot,  the scaling behavior and the scaling region is shown clearly. The scaling region is small as $ |1-N_f^c/N_f|\sim 10^{-3}$.  The scaling exponent is consistent with the mean field value $\Theta=2$~\cite{Bashir:2013zha}.

 The previous calculation identifies the critical point   at $N_f^c=6.81$. marking the disappearance  of the chiral symmetry breaking. However, it does not yet reach the conformal window {, since the gluon propagator still exhibits an infrared mass scale: the gluon dressing function in \cref{fig:gluon} shows a clear turnover at $p\sim m_g$, where $m_g^{-2}\equiv G_A(0)$.  
The mass scale gives a nonzero gluon condensate as suggested in Refs.~\cite{Cornwall:1981zr, Graziani:1984cs}, preventing  true  {conformal} behavior.   Notably,  we observe that  as the flavor number becomes larger, the gluon dressing function becomes flatter.  Since the gluon dressing represents  the running  of the coupling,  as illustrated by the effective charge or  the Taylor  coupling~\cite{Boucaud:2008gn,Aguilar:2010gm,Zafeiropoulos:2019flq},  the flattening  of the gluon dressing  serves as  the smoking gun for the  walking regime.  Besides, our results indicate that the walking regime is dominant by the gluon mass generation without chiral symmetry breaking. Such a pattern and the obtained $N_f^c$ for the walking regime  is  consistent with the previous results in DSE~\cite{Hopfer:2014zna}.
 
Crucially,  if  the scale of the theory in such a walking regime comes solely from the gluon condensate,   this  would support the symmetric mass generation that is proposed to explain the results for lattice QCD simulation with $N_f=8$~\cite{Hasenfratz:2022qan,Witzel:2025PoS146}.  Furthermore,   the symmetric mass generation is also applied to explain the emergence of  a new symmetry, i.e. the chiral spin symmetry,  observed  in the  QCD matter near above the chiral phase transition temperature~\cite{Glozman:2022lda}, where the phase is chiral symmetric but with nonzero gluon condensate.   It is therefore interesting to check the chiral spin symmetry among the mesons at $N_f=8$  to verify the connection  of  these phenomenons.

  It needs to mention that the current minimal QCD scheme employs the difference DSE for gluon propagator expanding on the quenched propagator, and  consequently,  it fails to track down the lower boundary of the conformal window, i.e. the location of the CBZ fixed point,   where the gluon condensate also vanishes. The incorporation  for  a full self-consistent determination of gluon and ghost propagator  can help to give an estimate for the CBZ fixed point, which we plan to pursue  in the future studies.

\begin{figure}[t]
\centering
\includegraphics[width=0.9\columnwidth]{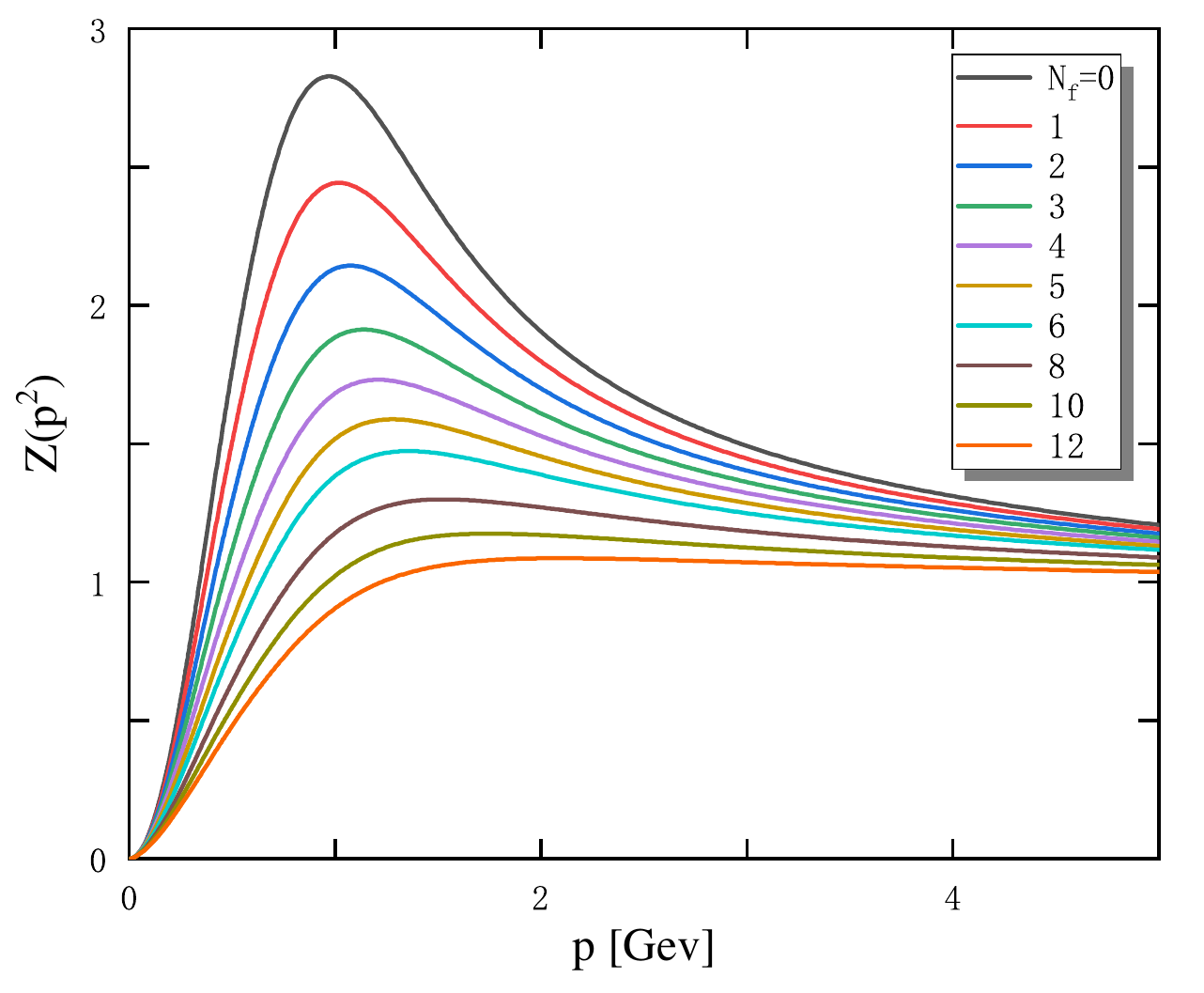} 
\caption{Gluon propagator dressing function as a function of flavor number. The flattening of the dressing function for large flavor number implies the walking regime.}
\label{fig:gluon}
\end{figure}

%%%%%%%%%%%%%%%%%%%%

%%%%%%%%%%%%%%%%%%%%%%%%%%%%%%%%%%%%%%
\section{Summary}
\label{sec:Summary}
This work presents a  non-perturbative study of the flavor dependence  for chiral symmetry breaking within QCD, employing the Dyson–Schwinger equation  approach. A minimal  QCD scheme is introduced, which maintains the running behavior of QCD quantitatively, in particular of the quark and gluon propagator.     The flavor dependence of the gluon dressing function and the quark mass function is investigated across 
$N_f\in [0,12]$. Our results reveal a phase transition marked by the disappearance of the chiral condensate at 
$N_f^c=6.81$, with a clear  scaling  behavior as $|N_f-N_f^c|^{0.53(9)}$.  Beyond this critical point, the flattening of the gluon dressing function indicates that the theory steps into the walking regime where the chiral symmetry is restored while  the gluon propagator retains a dynamical mass scale.  Limited by the current scheme of calculating the Yang-Mills sector, it is not capable of giving an estimate on the boundary of walking regime and conformal window.  Future refinements incorporating self-consistent gluon and ghost propagator calculations will help resolve this question. 
 
Now it is also interesting to  {consider} the mapping between the flavor dependence of QCD and  {its} phase structure at finite temperature. For instance, in QCD with eight massless flavors,  {a chiral phase transition at finite temperature is not expected,}  {since the theory already lies in the chiral symmetric walking regime at zero temperature with a vanishing chiral condensate}.  {Nevertheless}, there may exist a  {Miransky/BKT-type} transition{, or a crossover} between the  {walking} regime and the conformal limit which can be  {signaled} by the gluon condensate as discussed above. {For systems with fewer flavours},  this transition {may also take place at temperatures above the chiral transition,} establishing  a possible link  between the {walking} regime and the chiral spin symmetric matter  observed in the lattice QCD simulation. Therefore, a systematic  investigation  for the many flavor system at finite temperature  is in demand to  deepen the understanding of the walking regime and to uncover the properties of  QCD matter  near and above the chiral phase transition temperature.

\begin{acknowledgments}
 FG and YL thank J. Pawlowski and A. Pastor for very helpful discussions.
This work  is supported by the National Natural Science Foundation of China under Grants  No. 12305134   and No. 12247107,  No. 12175007 {and No. 12475105}. 
\end{acknowledgments}

\hfill 
\newpage

\bibliographystyle{unsrt}
\bibliography{references}

\end{document}